\documentclass[12pt]{amsart}
\advance\textheight\topskip
\allowdisplaybreaks

\usepackage{amsmath}
\usepackage{amsfonts,amssymb}
\usepackage[mathscr]{eucal}
\usepackage{upgreek}
\usepackage{afterpage}
\usepackage{graphicx}

\usepackage{times,mathptmx}

\usepackage[dvips,bookmarks=false,hyperfigures=false,a4paper,
  colorlinks=true, linkcolor=black, citecolor=black]{hyperref}

\numberwithin{equation}{section}
\makeatletter
\@addtoreset{equation}{section}
\@addtoreset{subsubsection}{section}

\def\@secnumfont{\bfseries}
\def\subsubsection{\@startsection{subsubsection}{3}%
  \z@{.5\linespacing\@plus.7\linespacing}{-.5em}%
  {\normalfont\bfseries}}
\def\paragraph{\@startsection{paragraph}{4}%
  \z@\z@{-\fontdimen2\font}%
  \normalfont\bfseries}
\def\subparagraph{\@startsection{subparagraph}{5}%
  \z@\z@{-\fontdimen2\font}%
  \normalfont\bfseries}

\makeatother

\newcommand{\chicolumn}{\boldsymbol{\upchi}}

\newcommand{\Kmine}{\mathsf{K}}
\newcommand{\Pmine}{\mathsf{P}}
\newcommand{\Smine}{\mathsf{S}}
\newcommand{\SSfull}{\boldsymbol{\mathscr{S}}}
\newcommand{\SSblock}{\mathbb{S}}

\newcommand{\zero}{\relax}

\newcommand{\voal}[1]{\mathscr{#1}} 

\newcommand{\WWW}{\voal{W}_{\pplus,\pminus}}

\newcommand{\pplus}{p}
\newcommand{\pminus}{p'}
\newcommand{\ppi}{(\pplus,\pminus)}

\newcommand{\setii}{\mathscr{I}_0}

\newcommand{\qgg}{\boldsymbol{\mathfrak{g}}} 

\newcommand{\rep}{\mathscr} 

\newcommand{\repX}{\rep{X}}

\newcommand{\repK}{\rep{K}}

\renewcommand{\geq}{\,{\geqslant}\,}
\renewcommand{\leq}{\,{\leqslant}\,}

\newcommand{\SLiiZ}{SL(2,\oZ)}

\newcommand{\oC}{\mathbb{C}}

\newcommand{\oZ}{\mathbb{Z}}

\newcommand{\mfrac}[2]{\raisebox{.8pt}{\mbox{\small$\displaystyle\frac{#1}{#2}$}}}
\newcommand{\ffrac}[2]{\raisebox{.5pt}{\mbox{\footnotesize$\displaystyle\frac{#1}{#2}$}}}
\newcommand{\fffrac}[2]{\raisebox{.9pt}{\mbox{\scriptsize$\displaystyle\frac{#1}{#2}$}}}
\newcommand{\half}{%
  \mathchoice{\ffrac{1}{2}}{\frac{1}{2}}{\frac{1}{2}}{\frac{1}{2}}}

\renewcommand{\tilde}{\widetilde}

\swapnumbers

\theoremstyle{definition}

\begin{document}

\title{A note on the logarithmic $(p,p')$ fusion}

\address{Lebedev Physics Institute \hfill\mbox{}\linebreak
  \texttt{ams@sci.lebedev.ru}}

\author[Semikhatov]{A.M.~Semikhatov}



\begin{abstract}
  The procedure in [Fuchs et al.] to obtain a fusion algebra from the
  modular transformation of characters in logarithmic conformal field
  models is extended to the $\ppi$ logarithmic models.  The resulting
  fusion algebra coincides with the Grothendieck ring of the quantum
  group of the $\ppi$ model.
\end{abstract}

\maketitle
\thispagestyle{empty}

\section{Introduction}
This paper is a remark on fusion in a class of logarithmic models of
conformal field theory~\cite{[Gurarie],[GK-rat],[G-alg]}.  In rational
conformal field models, fusion is related to modular transformations
of characters by the celebrated Verlinde formula~\cite{[V],[V+]}.
Because the Verlinde formula relies on the fact that the fusion
algebra is semisimple, it does not immediately extend to logarithmic
conformal field theories, where fusion algebras (starting with the
pioneering results in~\cite{[GK-fusion]}) are typically nonsemisimple.
The known extensions of the Verlinde formula to the nonsemisimple
realm rely on some extra input, in one form or another~\cite{[FK]}
(also see~\cite{[GR2]}).  In the prescription proposed
in~\cite{[FHST]}, this extra input can be related to a quantum-group
formulation.

The role of quantum groups in logarithmic conformal field theory
gradually emerged in~\cite{[FGST],[FGST2],[FGST3],[FGST-q]}
(see~\cite{[S-q]} for a summary and~\cite{[MT]} for some further
development), leading to a version of the Kazhdan--Lusztig ``duality''
between the extended algebra $W$ in a logarithmic conformal field
model and the corresponding quantum group $\qgg$.\footnote{These are
  factorizable ribbon quantum groups at even roots of unity;
  see~\cite{[FHT]} for their other use and~\cite{[KSSB]} for an
  interesting precursor of their occurrence in logarithmic models: the
  ribbon structure, the (co)integral, and the $M$[onodromy] matrix
  (cf.~\cite{[FGST],[FGST-q]}) are already present in~\cite{[KSSB]},
  albeit in a somewhat simpler situation.}  The most remarkable result
related to the Kazhdan--Lusztig duality is the coincidence of modular
group representations (the one generated from the $W$ characters and
the one carried by the center of~$\qgg$); also, the Grothendieck ring
of $\qgg$ is a natural candidate for the fusion algebra of
$W$-representations (we speak of the $K_0$-type fusion,
see~\cite{[FHST],[Fuchs]}).

For the $(p,1)$ logarithmic models, in particular, this
``quantum-group candidate fusion'' coincides with the fusion derived
in~\cite{[FHST]} from the characters, thus lending additional support
to the procedure proposed in~\cite{[FHST]}.  The aim of this paper is
to extend the existing state of consistency to $\ppi$ logarithmic
models: we propose a prescription whereby the modular transformations
of the characters of the extended algebra in the $\ppi$ logarithmic
model~\cite{[FGST3]} are converted into a nonsemisimple fusion
algebra, which turns out to coincide with the Grothendieck ring of the
corresponding quantum group $\qgg$~\cite{[FGST-q]}.  For this, we
follow the approach in~\cite{[FHST]} (also see~\cite{[FK]}) very
closely.  In Sec.~\ref{sec:modular}, we describe our starting point,
the modular group representation generated from the characters of the
extended algebra of the $\ppi$ logarithmic models.  In
Sec.~\ref{sec:verlinde}, we formulate the procedure to convert these
modular transformations to the following fusion algebra on $2p p'$
elements~$\repK^{\pm}_{r,r'}$ \cite{[FGST-q]}:
\begin{equation}\label{the-fusion}
  \repK^{\alpha}_{r,{r'}}\repK^{\beta}_{{s},s'}
  =\sum_{\substack{u=|r - {s}| + 1\\
      \mathrm{step}=2}}^{r + {s} - 1}
  \sum_{\substack{u'=|{r'} - s'| + 1\\
      \mathrm{step}=2}}^{{r'} + s' - 1}
  {\tilde{\repK}}^{\alpha\beta}_{u,u'},
\end{equation}
where $\alpha,\beta=\pm1$ and
\begin{equation*}
  {\tilde{\repK}}^{\alpha}_{r,{r'}} =
  \begin{cases}
    \repK^{\alpha}_{r,{r'}},& 
      1\leq r\leq \pplus,\quad
      1\leq {r'}\leq \pminus,
    \\[2pt]
    \repK^{\alpha}_{2\pplus - r,{r'}} + 2\repK^{-\alpha}_{r - \pplus, {r'}},&
      \pplus\!+\!1\leq r\leq 2 \pplus\!-\!1,\quad
      1\leq {r'}\leq \pminus,
    \\[2pt]
    \repK^{\alpha}_{r,2\pminus - {r'}} + 2\repK^{-\alpha}_{r,{r'} - \pminus},&
      1\leq r\leq \pplus,\quad
      \pminus\!+\!1\leq {r'}\leq 2 \pminus\!-\!1,
    \\[2pt]
    \mbox{}\kern-3pt\begin{aligned}[b]
      &\repK^{\alpha}_{2\pplus - r, 2 \pminus - {r'}}
      + 2\repK^{-\alpha}_{2\pplus - r, {r'} - \pminus}\\
      &{}+ 2\repK^{-\alpha}_{r - \pplus, 2 \pminus - {r'}}
      + 4\repK^{\alpha}_{r - \pplus, {r'} - \pminus},
    \end{aligned}&
      \pplus\!+\!1\leq r\leq 2 \pplus\!-\!1,\quad
      \pminus\!+\!1\leq {r'}\leq 2 \pminus\!-\!1.
  \end{cases}
\end{equation*}
The identity of this associative commutative algebra is given by
$\repK^+_{1,1}$.  We also recall from~\cite{[FGST-q]} that this
algebra is generated by two elements $\repK^+_{1,2}$ and
$\repK^+_{2,1}$ and can also be described as the quotient of \
$\oC[x,y]$ by the ideal generated by the polynomials
\begin{align*}
  &U_{2\pplus + 1}(x) - U_{2\pplus - 1}(x) - 2,\\
  &U_{2\pminus + 1}(y) - U_{2\pminus - 1}(y) - 2,\\
  &U_{\pplus + 1}(x) - U_{\pplus - 1}(x)
  -U_{\pminus + 1}(y) + U_{\pminus - 1}(y),
\end{align*}
where
\begin{gather*}
  U_s(2\cos t)=\mfrac{\sin s t}{\sin t},\quad s\geq1,
\end{gather*}
are Chebyshev polynomials of the second kind.

\section{Modular transformations of the $\ppi$
  characters~\cite{[FGST3]}}\label{sec:modular}
For each pair of coprime positive integers $p, p'$, the extended
algebra of the logarithmic $\ppi$ model is the $W$-algebra $\WWW$
identified and studied in~\cite{[FGST3]}.  It has $\half(p-1)(p'-1)+2p
p'$ irreducible representations, the $\half(p-1)(p'-1)$ of which are
just the Virasoro representations in the corresponding $\ppi$ minimal
model and the other are ``genuine'' $\WWW$-representations (such that
the radical of $\WWW$ acts nontrivially).  In what follows, the
characters of irreducible $\WWW$-representations are
denoted~as
\begin{equation}\label{the-chars}
  \underset{
    (r,r')\in\setii
  }{\chi_{r,r'}(\tau),}\quad
  \underset{\substack{1\leq r\leq p,\\
      1\leq r'\leq p'}}{\chi^+_{r,r'}(\tau),\ \ \chi^-_{r,r'}(\tau),}
\end{equation}
where we introduce the index set
\begin{equation}
  \label{eq:setii}
  \setii =\bigl\{(r,{r'}) \bigm| 1\leq r\leq \pplus\!-\!1,\ 1\leq
  {r'}\leq \pminus\!-\!1,\
  \pminus r+ \pplus {r'}\leq \pplus\pminus\bigr\},\pagebreak[3]
\end{equation}
with $|\setii|=\half(\pplus\,{-}\,1)(\pminus\,{-}\,1)$ (we recall the
well-known symmetry $\chi_{r,r'}(\tau)=\chi_{p-r,p'-r'}(\tau)$ of the
minimal-model Virasoro characters).

The modular (specifically, $S$-) transformation properties of the
characters are as follows.  First, the minimal-model characters
$\chi_{r,{r'}}$ are well-known to $S$-transform as
\begin{equation}\label{S-chi}
  \chi_{r,{r'}}(-\ffrac{1}{\tau})=
  {}-\ffrac{2\sqrt{2}}{\sqrt{\mathstrut \pplus\pminus}}
  \sum_{({s},s')\in\setii}\!\!
  (-1)^{r s' + {s} {r'}}\sin\ffrac{\pi \pminus r {s}}{\pplus}
  \,\sin\!\ffrac{\pi \pplus {r'} s'\!}{\pminus}\,\chi_{{s},s'}(\tau),
  \quad(r,{r'})\in\setii.
\end{equation}
Next, it follows from~\cite{[FGST3]} 
that (for $1\leq r\leq p$ and $1\leq r'\leq p'$)
\begin{multline}\label{S-chiP}
  \chi^+_{r, r'}(-\ffrac{1}{\tau}) = 
  \sum_{s=1}^{p}\smash{\sum_{s'=1}^{p'}}
  \mathscr{S}_{r, r'; s, s'}(\tau)\bigl(\chi^+_{s, s'}(\tau)
  + (-1)^{p' r + p r'}\chi^-_{s,s'}(\tau)\bigr)
  \\*[-6pt]
  {}+\sum_{(s,s')\in\setii}\!\!\!
  \tilde{\mathscr{S}}^+_{r,r';s,s'}(\tau)\chi_{s,s'}(\tau),
\end{multline}
\mbox{}\vspace*{-\baselineskip}
\begin{multline}\label{S-chiM}
  \chi^-_{r, r'}(-\ffrac{1}{\tau})= 
  \sum_{s=1}^{p}\sum_{s'=1}^{p'}
  (-1)^{p s' + p' s}\mathscr{S}_{r, r'; s, s'}(\tau)\bigl(\chi^+_{s, s'}(\tau)
  + (-1)^{p' r + p r'}\chi^-_{s,s'}(\tau)\bigr)
  \\[-6pt]
  {}+\sum_{(s,s')\in\setii}\!\!\!
  \tilde{\mathscr{S}}^-_{r,r';s,s'}(\tau)\chi_{s,s'}(\tau),
\end{multline}
where the matrix elements $\mathscr{S}_{r, r', s, s'}(\tau)$ that
interest us in what follows are given by
\begin{equation}\label{the-S}
  \kern-5pt\begin{alignedat}{2}
    \mathscr{S}_{r, r'; s, s'}(\tau)&=
    \fffrac{2\sqrt{2}}{\sqrt{p p'}}(-1)^{r s' + s r'}
    \bigl(\ffrac{r}{p}\,\cos\!\ffrac{\pi p' r s\!}{p}
    - i\tau\ffrac{p\!-\!s}{p}\,\sin\!\ffrac{\pi p' r s\!}{p}\bigr)\\
    &\qquad\qquad\qquad\qquad{}\times
    \bigl(\ffrac{r'}{p'}\,\cos\!\ffrac{\pi p r' s'\!}{p'}
    - i\tau\ffrac{p'\!-\!s'}{p'}\,\sin\!\ffrac{\pi p r' s'\!}{p'}\bigr),&
    &\smash[t]{\kern-4pt\begin{array}[b]{l}
        1\leq s\leq p\!-\!1,\\
        1\leq s'\leq p'\!-\!1,
      \end{array}}
    \\  
    \mathscr{S}_{r, r'; s, p'}(\tau)&=
    \fffrac{\sqrt{2}}{\sqrt{p p'}}\,\ffrac{r'}{p'}
    (-1)^{s r' + p r' + p' r}
    \bigl(\ffrac{r}{p}\,\cos\!\ffrac{\pi p' r s\!}{p}
    - i\tau\ffrac{p\!-\!s}{p}\,\sin\!\ffrac{\pi p' r s\!}{p}\bigr),
    &\ &1\leq s\leq p\!-\!1
    \\
    \mathscr{S}_{r, r'; p, s'}(\tau)&=
    \fffrac{\sqrt{2}}{\sqrt{p p'}}\,\ffrac{r}{p}
    (-1)^{s' r + p' r + p r'}
    \bigl(\ffrac{r'}{p'}\,\cos\!\ffrac{\pi p r' s'\!}{p'}
    - i\tau\ffrac{p'\!-\!s'}{p'}\,\sin\!\ffrac{\pi p r' s'\!}{p'}\bigr),
    &\ &1\leq s'\leq p'\!-\!1
    \\
    \mathscr{S}_{r, r'; p, p'}(\tau)&= \fffrac{1}{\sqrt{2 p p'}}\,
    \ffrac{r r'}{p p'},
  \end{alignedat}\kern-15pt
\end{equation}
and the other matrix elements are
\begin{small}%
\begin{multline*}
  \tilde{\mathscr{S}}^+_{r,r';s,s'}(\tau)
  =(-1)^{r s' + s r'}\fffrac{\sqrt{2}}{p^2 p'^2\sqrt{p p'}}
  \Bigl(
  p p' r r'\,
  \cos\!\fffrac{\pi p' r s\!}{p}\,\cos\!\fffrac{\pi p r' s'\!}{p'}
  \\
   + i p' r   \tau(p s' - p' s)\,
   \cos\!\fffrac{\pi p' r s\!}{p}\,\sin\!\fffrac{\pi p r' s'\!}{p'}
  + i p r' \tau(p' s - p s')\,
  \sin\!\fffrac{\pi p' r s\!}{p}\,\cos\!\fffrac{\pi p r' s'\!}{p'}
  \\
  +\bigl(\fffrac{(p s' - p' s)^2}{2}\,\tau^2 - 2 i \pi p p' \tau
  +\fffrac{p^2 r'^2 + p'^2 r^2}{2}\bigr)
  \sin\!\fffrac{\pi p' r s\!}{p}\,\sin\!\fffrac{\pi p r' s'\!}{p'}
  \Bigr),
\end{multline*}
\mbox{}\vspace*{-1.2\baselineskip}
\begin{multline*}
  \ \tilde{\mathscr{S}}^-_{r,r';s,s'}(\tau)
  =(-1)^{s p' + s' p}\tilde{\mathscr{S}}^+_{r,r';s,s'}(\tau)
  -(-1)^{r s' + s r' + s p' + s' p}
  \fffrac{1}{\sqrt{2 p p'}}\,
  \sin\!\fffrac{\pi p' r s\!}{p}\,
  \sin\!\fffrac{\pi p r' s'\!}{p'}.\hfill
\end{multline*}%
\end{small}%

In~\eqref{the-S}, remarkably, the dependence on the primed and
unprimed indices almost (modulo $(-1)^{r s' + s r'}$) factors, which
\textit{partly} reduces the analysis to that for the $(p,1)$ and
$(1,p')$ cases.  (Most of the quantum-group objects corresponding to
the $(p,p')$ models in~\cite{[FGST-q]} also have an ``almost
factored'' form.)

\section{``Logarithmic'' $\ppi$-fusion}\label{sec:verlinde}
\subsection{The procedure}
The steps leading from~\eqref{S-chiP} and~\eqref{S-chiM}
to~\eqref{the-fusion}, in much the same way as in~\cite{[FHST]}, are
as follows.

\begin{enumerate}\addtolength{\itemsep}{4pt}%
\renewcommand{\labelenumi}{\theenumi.}
\item We view the characters in~\eqref{the-chars} as a column vector
  and write the $S$-transformation formulas as
  \begin{equation*}
    \chicolumn(-\ffrac{1}{\tau})
    =\SSfull(\tau)\chicolumn(\tau), 
  \end{equation*}
  with the corresponding $N\times N$ $\tau$-dependent matrix
  $\SSfull(\tau)$, where
  \begin{equation*}
      N=\half(\pplus\,{-}\,1)(\pminus\,{-}\,1)+2\pplus \pminus
  \end{equation*}
  is the total number of characters.
  
  We then take $\SSblock(\tau)$ to be the $(2\pplus \pminus)\times
  (2\pplus \pminus)$ block of $\SSfull(\tau)$ corresponding to the
  $2\pplus \pminus$ characters $\chi^{\pm}_{r,{r'}}(\tau)$, $1\leq
  r\leq \pplus$, $1\leq {r'}\leq \pminus$.  That is, we deal with only
  the $\mathscr{S}_{r, r'; s, s'}(\tau)$ in~\eqref{the-S}.

  In accordance with the block structure of the Jordan form of
  $\SSblock(\tau)$, we fix the block structure of matrices as follows:
  $2$ blocks of size $1\times 1$, \
  $(\pplus\,{-}\,1)+(\pminus\,{-}\,1)$ blocks of size $2\times 2$, and
  $\half(\pplus\,{-}\,1)(\pminus\,{-}\,1)$ blocks of size $4\times 4$.
  
\item Totally similarly to~\cite{[FHST]}, there exists a ($(2\pplus
  \pminus)\times (2\pplus \pminus)$-matrix) automorphy factor
  $J(\gamma,\tau)$, with $\gamma\in\SLiiZ$, satisfying the cocycle
  condition and a commutativity property formulated in~\cite{[FHST]},
  such that
  \begin{gather*}
    \Smine=J(S, \tau)\SSblock(\tau)
  \end{gather*}
  is a numerical ($\tau$-independent) matrix; in fact,
  \begin{equation}\label{mS}
    \Smine=\SSblock(i).
  \end{equation}
  It then follows that $\Smine^2=1$.\enlargethispage{\baselineskip}

\item From now on, $\chicolumn=(\chicolumn_J)$ denotes the $2\pplus
  \pminus$ $\WWW$-characters ordered as
  \begin{equation}\label{ordering}
    \chicolumn=
    (\chi^+_{\pplus,\pminus},\chi^-_{\pplus,\pminus},
    \underbrace{\chi^+_{r,\pminus},\chi^-_{\pplus-r,\pminus}}_{1\leq r\leq \pplus-1},
    \underbrace{\chi^+_{\pplus,{r'}},\chi^-_{\pplus,\pminus-{r'}}}_{1\leq {r'}\leq \pminus-1},
    \underbrace{\chi^+_{r,{r'}},\chi^-_{\pplus-r,{r'}},
    \chi^-_{r,\pminus-{r'}},\chi^+_{\pplus-r,\pminus-{r'}}}_{(r,{r'})\in\setii}).
  \end{equation}  
  This arrangement of the characters clearly agrees with the
  ``$1+1+(\pplus\,{+}\,\pminus\,{-}\,2)\cdot 2\times 2+
  \half(\pplus\,{-}\,1)(\pminus\,{-}\,1)\cdot 4\times 4$'' block
  structure.  We also define a special row\pagebreak[3]
  \begin{equation*}
    \Pmine_\Omega=
    (1,1,
    \underbrace{1,0,\,\dots,\,1,0}_{2(\pplus-1)\ \text{elements}},
    \underbrace{1,0,\,\dots,\,1,0}_{2(\pminus-1)\ \text{elements}},
    \underbrace{1,0,0,0,\,\dots,\,1,0,0,0}_{4\cdot\half(\pplus-1)(\pminus-1)\
      \text{elements}}).
  \end{equation*}
  (At this point, we anticipate that the block structure inherited
  from $\SSblock(\tau)$ is to become the block structure of the fusion
  algebra; accordingly, the ones encountered in $\Pmine_\Omega$
  correspond to the decomposition of the identity into a sum of
  primitive idempotents, one for each block.)

\item Let $\Smine_\Omega=(\Smine_\Omega^{\ \ J})$ be the row of
  $\Smine$ corresponding to the vacuum-representation character
  $\chicolumn_\Omega=\chi^+_{1,1}$,\footnote{The vacuum representation
    $\repK^+_{1,1}$ of the $\WWW$ algebra in~\cite{[FGST3]} is in fact
    an extension of the Virasoro representation $\repX_{1,1}$ whose
    character is $\chi_{1,1}(\tau)$ by the $\WWW$-representation
    $\repX^+_{1,1}$ whose character is $\chi^+_{1,1}(\tau)$:
    $0\to\repX^+_{1,1}\to\repK^+_{1,1}\to\repX_{1,1}\to0$.  The
    difference between $\repK^+_{1,1}$ and $\repX^+_{1,1}$ is
    irrelevant in the present context, where we ignore all the
    $\chi_{r,r'}$ characters altogether.} i.e.,
  \begin{equation*}
    \chi^+_{1,1}(-\ffrac{1}{\tau})=\Smine_\Omega^{\ \ J}\chicolumn_J(\tau)
  \end{equation*}
  (the sum is taken over the $2\pplus \pminus$ values of $J$ in
  accordance with~\eqref{ordering}).  With the chosen ordering,
  $\chicolumn_\Omega$ occupies position $2\pplus\!+\!2\pminus\!-\!1$
  in~\eqref{ordering} and, accordingly, $\Smine_\Omega$ is the
  $(2\pplus\!+\!2\pminus\!-\!1)$th row.  Explicitly
  (see~\eqref{the-S}), the segment of $\Smine_\Omega$ corresponding to
  $(\chi^+_{s,{s'}},\chi^-_{\pplus-s,{s'}},
  \chi^-_{s,\pminus-{s'}},$\linebreak[0]$\chi^+_{\pplus-s,\pminus-{s'}})$
  is given by $\fffrac{2\sqrt{2}}{p p'\sqrt{p p'}}(-1)^{s' + s}$ times
  \begin{multline}\label{S-segment}
    \Bigl(\!
    \bigl(\cos\!\ffrac{\pi p' s\!}{p}
    + (p\!-\!s)\sin\!\ffrac{\pi p' s\!}{p}\bigr)
    \bigl(\cos\!\ffrac{\pi p s'\!}{p'}
    + (p'\!-\!s')\sin\!\ffrac{\pi p s'\!}{p'}\bigr),
    \\
    \bigl(\cos\!\ffrac{\pi p' s\!}{p}
    - s\,\sin\!\ffrac{\pi p' s\!}{p}\bigr)
    \bigl(\cos\!\ffrac{\pi p s'\!}{p'}
    + (p'\!-\!s')\sin\!\ffrac{\pi p s'\!}{p'}\bigr),
    \qquad
    \\
    \qquad
    \bigl(\cos\!\ffrac{\pi p' s\!}{p}
    + (p\!-\!s)\sin\!\ffrac{\pi p' s\!}{p}\bigr)
    \bigl(\cos\!\ffrac{\pi p s'\!}{p'}
    - s'\,\sin\!\ffrac{\pi p s'\!}{p'}\bigr),
    \\
    \bigl(\cos\!\ffrac{\pi p' s\!}{p}
    - s\,\sin\!\ffrac{\pi p' s\!}{p}\bigr)
    \bigl(\cos\!\ffrac{\pi p s'\!}{p'}
    - s'\,\sin\!\ffrac{\pi p s'\!}{p'}\bigr)
    \!\Bigr).
  \end{multline}
  
\item We next consider the equation (cf.~\cite{[FHST]})
  \begin{equation}\label{P=SK}
    \Pmine_\Omega=\Smine_\Omega \Kmine
  \end{equation}
  and solve it for the block-diagonal matrix
  \begin{equation}\label{The-K}
    \renewcommand{\arraycolsep}{0pt}
    \Kmine=
    \mbox{\footnotesize$\displaystyle
    \begin{pmatrix}
      \kappa_1\\
      {}&\kappa_2\\
      {}&{}&\boxed{K_{2\times2}\rule[-4pt]{0pt}{13pt}\!}\\[-4pt]
      {}&{}&{}&\ddots\\[-2pt]
      {}&{}&{}&{}&{}\boxed{K_{2\times2}\rule[-4pt]{0pt}{13pt}\!}\\
      {}&{}&{}&{}&{}&{}\boxed{K_{4\times4}\rule[-4pt]{0pt}{13pt}\!}\\[-4pt]
      {}&{}&{}&{}&{}&{}&\ddots\\[-2pt]
      {}&{}&{}&{}&{}&{}&{}&{}\boxed{K_{4\times4}\rule[-4pt]{0pt}{13pt}\!}
    \end{pmatrix}$}
  \end{equation}
  (with zeros outside the blocks), where
  the $2\times 2$ blocks are as in~\cite{[FHST]}, i.e., have the
  structure
  \begin{equation*}
    K_{2\times2}=\begin{pmatrix}
      a&\lambda\\
      -a&b\lambda
    \end{pmatrix}
  \end{equation*}
  (it is understood that
  $K^{(i)}_{2\times2}=\mbox{\footnotesize$\begin{pmatrix}
      a^{(i)}\!\!&\!\!\!\lambda^{(i)}\\[-2pt]
      -a^{(i)}\!\!&\!\!\!b^{(i)}\lambda^{(i)}
    \end{pmatrix}$}$ for each block, but the block dependence is
  omitted for brevity) and the $4\times4$ blocks have the structure
\begin{equation*}
    K_{4\times4}=\begin{pmatrix}
      a& \mu& \nu&  \fffrac{1}{a}\,\mu \nu&\\[6pt]
      -a& -\mu& c\nu& \ffrac{c}{a}\,\mu \nu&\\[6pt]
      -a& b\mu& -\nu& \fffrac{b}{a}\,\mu \nu\\[6pt]
      a& -b\mu& -c\nu& \fffrac{b c}{a}\,\mu \nu
    \end{pmatrix}
  \end{equation*}
  (again, with the block dependence omitted).

  The nonzero factors $\lambda$, $\mu$, and $\nu$, rescaling each
  column except the first in each block, are irrelevant in what
  follows (because nilpotent elements have no canonical
  normalization).  The unknowns $a$ and $b$ in each $2\times2$ block
  and $a$, $b$, and $c$ in each $4\times4$ block are determined from
  Eq.~\eqref{P=SK}.  That is, if $(s_1,s_2,s_3,s_4)$ is a segment of
  $\Smine_\Omega$ corresponding to a $4\times4$ block, then
  \begin{equation*}
    a=\ffrac{1}{s_1 - s_2 - s_3 + s_4},
    \quad
    b=\ffrac{s_2 - s_1}{s_3 - s_4},\quad
    c=\ffrac{s_3 - s_1}{s_2 - s_4}
  \end{equation*}
  in this block; the equations are compatible because $s_1 s_4=s_2
  s_3$, as is readily seen from~\eqref{S-segment}.  (By~\eqref{P=SK},
  the two elements of $\Kmine$ that constitute the $1\times1$ blocks
  are the inverse of the corresponding $\Smine$-matrix coefficients,
  just as the denominators in the semisimple Verlinde formula; with
  the $2\times2$ blocks, the situation repeats that in~\cite{[FHST]}.)
  
\item We set
  \begin{equation*}
    \Pmine=\Smine\Kmine.
  \end{equation*}
  
  The fusion algebra is now reconstructed from the $\Pmine$ matrix in
  much the same way as in~\cite{[FHST]}, as follows.  Evidently, the
  $(2\pplus + 2\pminus -1)$th row of~$\Pmine$ is just $\Pmine_\Omega$.
  Let $\Pmine_I$ be the $I$th row of $\Pmine$.  We define
  $\mathsf{M}_I$, $I=1,\dots,2\pplus \pminus$, to be block-diagonal
  matrices that solve the equation
  \begin{equation}\label{pi-M}
    \Pmine_I=\Pmine_{\Omega}\,\mathsf{M}_I
  \end{equation}
  and whose $2\times2$ blocks are of the form (just as
  in~\cite{[FHST]})
  \begin{equation*}
    \begin{pmatrix}
      \alpha&\beta\\
      0&\alpha
    \end{pmatrix}
  \end{equation*}
  and the $4\times4$ blocks are
  \begin{equation*}
    \begin{pmatrix}
      \alpha&\beta&\gamma&\zeta\\
      {}&\alpha&0&\gamma\\
      {}&{}&\alpha&\beta\\
      {}&{}&{}&\alpha
    \end{pmatrix}
  \end{equation*}
  (with zeros below the diagonal).
  
  The $\mathsf{M}_I$ are then determined \textit{uniquely}; in
  particular, the $4\times4$ blocks are given by
  \begin{equation*}
    \begin{pmatrix}
      p_I&q_I&r_I&s_I\\
      {}&p_I&0&r_I\\
      {}&{}&p_I&q_I\\
      {}&{}&{}&p_I
    \end{pmatrix},
  \end{equation*}
  where $(p_I, q_I, r_I, s_I)$ is a segment of $\Pmine_I$
  corresponding to the chosen block.
\end{enumerate}

The result is then that the $\mathsf{M}_I$ satisfy the algebra
\begin{equation}\label{MMM}
  \mathsf{M}_I\mathsf{M}_J=\sum_K n_{IJ}^K \mathsf{M}_K,
\end{equation}
where the nonnegative integer coefficients $n_{IJ}^K$ turn out to be
those read off from \eqref{the-fusion}.  (Simultaneously, the matrix
$\mathsf{N}_I=\Pmine\mathsf{M}_I\,\Pmine^{-1}$ for each $I$ gives the
fusion structure constants as $(\mathsf{N}_I)_J^{\ K} =n_{IJ}^{K}$.)
Evidently, $\mathsf{M}_\Omega$ is the unit in the algebra.

\subsection{Examples}
The illustrative power of examples is hampered by the rapidly growing
matrix size and the general clumsiness of explicit expressions.  We
consider only the ``percolation'' and ``Lee--Yang'' cases, where
explicit values of the various matrix entries may be useful for
comparison with the studies of these cases by more direct methods
(e.g., in~\cite{[EF]}).

\subsubsection{$(3,2)$}\label{app:32-example}
For $\ppi=(3,2)$, the $12\times 12$ matrix $\Smine=\SSblock(i)$
explicitly evaluates as
\begin{equation*}
  \Smine=\mbox{\footnotesize$\displaystyle
    \setcounter{MaxMatrixCols}{12}
    \renewcommand{\arraycolsep}{2pt}
    \begin{pmatrix}
      \frac{1}{2{\sqrt{3}}}& 
      \frac{1}{2{\sqrt{3}}} & \frac{1}
      {{\sqrt{3}}} & \frac{1}
      {{\sqrt{3}}} & \frac{1}
      {{\sqrt{3}}} & \frac{1}
      {{\sqrt{3}}} & \frac{1}
      {{\sqrt{3}}} & \frac{1}
      {{\sqrt{3}}} & \frac{2}
      {{\sqrt{3}}} & \frac{2}
      {{\sqrt{3}}} & \frac{2}
      {{\sqrt{3}}} & \frac{2}
      {{\sqrt{3}}} \\ \frac{1}
      {2{\sqrt{3}}} & \frac{1}
      {2{\sqrt{3}}} & \frac{1}
      {{\sqrt{3}}} & \frac{1}
      {{\sqrt{3}}} & \frac{1}
      {{\sqrt{3}}} & \frac{1}
      {{\sqrt{3}}} & \frac{-1}
      {{\sqrt{3}}}   & \frac{-1}{{\sqrt{3}}}  &
      \frac{-2}{{\sqrt{3}}} & \frac{-2}
      {{\sqrt{3}}} & \frac{-2}
      {{\sqrt{3}}} & \frac{-2}
      {{\sqrt{3}}} \\ \frac{1}
      {6{\sqrt{3}}} & \frac{1}
      {6{\sqrt{3}}} & \frac{6 - 
        {\sqrt{3}}}{18} & \frac{-3 - 
        {\sqrt{3}}}{18} & \frac{-3 - 
        {\sqrt{3}}}{18} & \frac{6 - 
        {\sqrt{3}}}{18} & \frac{1}
      {3{\sqrt{3}}} & \frac{1}
      {3{\sqrt{3}}} & \frac{6 - 
        {\sqrt{3}}}{9} & \frac{-3 - 
        {\sqrt{3}}}{9} & \frac{6 - 
        {\sqrt{3}}}{9} & \frac{-3 - 
        {\sqrt{3}}}{9} \\ \frac{1}
      {3{\sqrt{3}}} & \frac{1}
      {3{\sqrt{3}}} & \frac{-3 - 
        {\sqrt{3}}}{9} & \frac{3 - 
        2{\sqrt{3}}}{18} & \frac{3 - 
        2{\sqrt{3}}}{18} & \frac{-3 - 
        {\sqrt{3}}}{9} & \frac{2}
      {3{\sqrt{3}}} & \frac{2}
      {3{\sqrt{3}}} & \frac{-2
        ( 3 + {\sqrt{3}} ) }{9}
      & \frac{3 - 2{\sqrt{3}}}{9} & 
      \frac{-2( 3 + {\sqrt{3}}
        ) }{9} & \frac{3 - 
        2{\sqrt{3}}}{9} \\ \frac{1}
      {3{\sqrt{3}}} & \frac{1}
      {3{\sqrt{3}}} & \frac{-3 - 
        {\sqrt{3}}}{9} & \frac{3 - 
        2{\sqrt{3}}}{18} & \frac{3 - 
        2{\sqrt{3}}}{18} & \frac{-3 - 
        {\sqrt{3}}}{9} & \frac{-2}
      {3{\sqrt{3}}} & \frac{-2}
      {3{\sqrt{3}}} & \frac{2
        ( 3 + {\sqrt{3}} ) }{9}
      & \frac{2{\sqrt{3}} - 3}
      {9} & \frac{2
        ( 3 + {\sqrt{3}} ) }{9}
      & \frac{2{\sqrt{3}} - 3}
      {9} \\ \frac{1}
      {6{\sqrt{3}}} & \frac{1}
      {6{\sqrt{3}}} & \frac{6 - 
        {\sqrt{3}}}{18} & \frac{-3 - 
        {\sqrt{3}}}{18} & \frac{-3 - 
        {\sqrt{3}}}{18} & \frac{6 - 
        {\sqrt{3}}}{18} & \frac{-1}
      {3{\sqrt{3}}} & \frac{-1}
      {3{\sqrt{3}}} & \frac{{\sqrt{3}} - 6}{9} & \frac{3 + 
        {\sqrt{3}}}{9} & \frac{{\sqrt{3}} - 6}{9} & \frac{3 + 
        {\sqrt{3}}}{9} \\ \frac{1}
      {4{\sqrt{3}}} & \frac{-1}
      {4{\sqrt{3}}} & \frac{1}
      {2{\sqrt{3}}} & \frac{1}
      {2{\sqrt{3}}} & \frac{-1}
      {2{\sqrt{3}}} & \frac{-1}
      {2{\sqrt{3}}} & \frac{-1}
      {2{\sqrt{3}}} & \frac{1}
      {2{\sqrt{3}}} & \frac{-1}
      {{\sqrt{3}}}   & -
      \frac{1}{{\sqrt{3}}}  &
      \frac{1}{{\sqrt{3}}} & \frac{1}
      {{\sqrt{3}}} \\ \frac{1}
      {4{\sqrt{3}}} & \frac{-1}
      {4{\sqrt{3}}} & \frac{1}
      {2{\sqrt{3}}} & \frac{1}
      {2{\sqrt{3}}} & \frac{-1}
      {2{\sqrt{3}}} & \frac{-1}
      {2{\sqrt{3}}} & \frac{1}
      {2{\sqrt{3}}} & \frac{-1}
      {2{\sqrt{3}}} & \frac{1}
      {{\sqrt{3}}} & \frac{1}
      {{\sqrt{3}}} & \frac{-1}
      {{\sqrt{3}}}   & -
      \frac{1}{{\sqrt{3}}}   \\
      \frac{1}{12{\sqrt{3}}} & \frac{-1}
      {12{\sqrt{3}}} & \frac{6 - 
        {\sqrt{3}}}{36} & \frac{-3 - 
        {\sqrt{3}}}{36} & \frac{3 + 
        {\sqrt{3}}}{36} & \frac{{\sqrt{3}} - 6}{36} & \frac{-1}
      {6{\sqrt{3}}} & \frac{1}
      {6{\sqrt{3}}} & \frac{{\sqrt{3}} - 6}{18} & \frac{3 + 
        {\sqrt{3}}}{18} & \frac{6 - 
        {\sqrt{3}}}{18} & \frac{-3 - 
        {\sqrt{3}}}{18} \\ \frac{1}
      {6{\sqrt{3}}} & \frac{-1}
      {6{\sqrt{3}}} & \frac{-3 - 
        {\sqrt{3}}}{18} & \frac{3 - 
        2{\sqrt{3}}}{36} & \frac{2{\sqrt{3}} - 3}{36} & \frac{3 + 
        {\sqrt{3}}}{18} & \frac{-1}
      {3{\sqrt{3}}} & \frac{1}
      {3{\sqrt{3}}} & \frac{3 + 
        {\sqrt{3}}}{9} & \frac{2{\sqrt{3}} - 3}{18} & \frac{-3 - 
        {\sqrt{3}}}{9} & \frac{3 - 
        2{\sqrt{3}}}{18} \\ \frac{1}
      {12{\sqrt{3}}} & \frac{-1}
      {12{\sqrt{3}}} & \frac{6 - 
        {\sqrt{3}}}{36} & \frac{-3 - 
        {\sqrt{3}}}{36} & \frac{3 + 
        {\sqrt{3}}}{36} & \frac{{\sqrt{3}} - 6}{36} & \frac{1}
      {6{\sqrt{3}}} & \frac{-1}
      {6{\sqrt{3}}} & \frac{6 - 
        {\sqrt{3}}}{18} & \frac{-3 - 
        {\sqrt{3}}}{18} & \frac{{\sqrt{3}} - 6}{18} & \frac{3 + 
        {\sqrt{3}}}{18} \\ \frac{1}
      {6{\sqrt{3}}} & \frac{-1}
      {6{\sqrt{3}}} & \frac{-3 - 
        {\sqrt{3}}}{18} & \frac{3 - 
        2{\sqrt{3}}}{36} & \frac{2{\sqrt{3}} - 3}{36} & \frac{3 + 
        {\sqrt{3}}}{18} & \frac{1}
      {3{\sqrt{3}}} & \frac{-1}
      {3{\sqrt{3}}} & \frac{-3 - 
        {\sqrt{3}}}{9} & \frac{3 - 
        2{\sqrt{3}}}{18} & \frac{3 + 
        {\sqrt{3}}}{9} & \frac{2{\sqrt{3}} - 3}{18}
    \end{pmatrix}.
    $}
\end{equation*}
Here, $\Smine_\Omega$ is the $9$th row of $\Smine$.  The matrix
$\Kmine$ in~\eqref{The-K} is then given
by
\begin{equation*}
  \Kmine=
  \mbox{\footnotesize$\displaystyle
    \setcounter{MaxMatrixCols}{20}
    \renewcommand{\arraycolsep}{3pt}
    \begin{pmatrix}
      12 {\sqrt{3}} & \zero & \zero & \zero & \zero & \zero & \zero &
      \zero & \zero & \zero & \zero & \zero \\ \zero & -12
      {\sqrt{3}} & \zero & \zero & \zero & \zero & \zero & \zero &
      \zero & \zero & \zero & \zero \\ \zero & \zero & 4 & 1 & 
      \zero & \zero & \zero & \zero & \zero & \zero & \zero & \zero \\ 
      \zero & \zero & -4 & \frac{7 - 3{\sqrt{3}}}
      {2} & \zero & \zero & \zero & \zero & \zero & \zero & \zero & 
      \zero \\ \zero & \zero & \zero & \zero & 4 & 1 & \zero & 
      \zero & \zero & \zero & \zero & \zero \\ \zero & \zero & \zero & 
      \zero & -4 & \frac{7 + 3{\sqrt{3}}}
      {11} & \zero & \zero & \zero & \zero & \zero & \zero \\ \zero &
      \zero & \zero & \zero & \zero & \zero & -3
      {\sqrt{3}} & 1 & \zero & \zero & \zero & \zero \\ 
      \zero & \zero & \zero & \zero & \zero & \zero & 3
      {\sqrt{3}} & 1 & \zero & \zero & \zero & \zero \\ 
      \zero & \zero & \zero & \zero & \zero & \zero & \zero & \zero & 
      -1 & 1 & 1 & 
      -1 \\ \zero & \zero & \zero & \zero & \zero & \zero & \zero & 
      \zero & 1 & -1 & \frac{7 - 3{\sqrt{3}}}
      {2} & \frac{3{\sqrt{3}} - 7}
      {2} \\ \zero & \zero & \zero & \zero & \zero & \zero & \zero & 
      \zero & 1 & 1 & -1 & 
      -1 \\ \zero & \zero & \zero & \zero & \zero & \zero & \zero & 
      \zero & -1 & -1 & \frac{3{\sqrt{3}} - 7}{2} &
      \frac{3{\sqrt{3}} - 7}{2}
    \end{pmatrix},$}
\end{equation*}
which gives rise to the fusion-algebra eigenmatrix
\begin{equation*}
  \Pmine=\Smine\Kmine=
  \mbox{\footnotesize$\displaystyle
    \setcounter{MaxMatrixCols}{13}
    \renewcommand{\arraycolsep}{2pt}
    \begin{pmatrix}
      6 & -6 & 0 & \frac{3
        ( \sqrt{3} - 1 ) }{2}
      & 0 & \frac{3
        ( 1 + 2{\sqrt{3}} ) }
      {11} & 0 & \frac{2}
      {{\sqrt{3}}} & 0 & 0 & 0 &
      6(1\,{-}\,{\sqrt{3}})
      \\
      6 & -6 & 0 & 
      \frac{3( \sqrt{3} - 1
        ) }{2} & 0 & \frac{3
        ( 1 + 2{\sqrt{3}} ) }
      {11} & 0 & \frac{-2}
      {{\sqrt{3}}} & 0 & 0 & 0 & 6
      ( \sqrt{3}\,{-}\,1 )  \\ 
      2 & -2 & 2 & 0 & -2 & 0 & 0 & 
      \frac{2}{3{\sqrt{3}}} & 0 & 2 & 
      0 & 0 \\ 4 & -4 & -2 & \frac{3
        ( 1 - \sqrt{3} ) }{4}
      & 2 & \frac{-3
        ( 1 + 2{\sqrt{3}} ) }
      {22} & 0 & \frac{4}
      {3{\sqrt{3}}} & 0 & -2 & 0 & 3
      ( \sqrt{3}\,{-}\,1 )
      \\ 
      4 & -4 & -2 & \frac{3
        ( 1 - \sqrt{3} ) }{4}
      & 2 & \frac{-3
        ( 1 + 2{\sqrt{3}} ) }
      {22} & 0 & \frac{-4}
      {3{\sqrt{3}}} & 0 & 2 & 0 &
      3(1\,{-}\,{\sqrt{3}}) \\
      2 & -2 & 2 & 0 & 
      -2 & 0 & 0 & \frac{-2}
      {3{\sqrt{3}}} & 0 & 
      -2 & 0 & 0 \\ 3 & 3 & 0 & 
      \frac{3( \sqrt{3} - 1
        ) }{4} & 0 & \frac{-3
        ( 1 + 2{\sqrt{3}} ) }
      {22} & 3 & 0 & 0 & 0 & 3(1\,{-}\,{\sqrt{3}}) & 0
      \\
      3 & 3 & 0 &
      \frac{3( \sqrt{3} - 1) }{4} & 0 & \frac{-3
        ( 1 + 2{\sqrt{3}} ) }
      {22} & -3 & 0 & 0 & 0 & 3
      ( \sqrt{3}\,{-}\,1 )  & 
      0 \\ 1 & 1 & 1 & 0 & 1 & 0 & 1 & 
      0 & 1 & 0 & 0 & 0
      \\
      2 & 2 & 
      -1 & \frac{3
        ( 1 - \sqrt{3} ) }{8}
      & -1 & \frac{3
        ( 1 + 2{\sqrt{3}} ) }
      {44} & 2 & 0 & -1 & 0 & \frac{3
        ( \sqrt{3} - 1 ) }{2}
      & 0 \\ 1 & 1 & 1 & 0 & 1 & 0 & 
      -1 & 0 & 
      -1 & 0 & 0 & 0 \\ 2 & 2 & 
      -1 & \frac{3
        ( 1 - \sqrt{3} ) }{8}
      & -1 & \frac{3
        ( 1 + 2{\sqrt{3}} ) }
      {44} & -2 & 0 & 1 & 0 & \frac{3
        ( 1 - \sqrt{3} ) }{2}
      & 0 \\  
    \end{pmatrix}$}
\end{equation*}
The fusion relations that follow in accordance
with~\eqref{pi-M}--\eqref{MMM} are the $(\pplus=3,\pminus=2)$
specialization of~\eqref{the-fusion} (explicitly written
in~\cite{[FGST3]}).

\subsubsection{$(5,2)$} 
For $\ppi=(5,2)$, all of the entries of the $20\times20$ matrix
$\Smine$ can be easily evaluated from the $\mathscr{S}_{r, r'; s,
  s'}(i)$ in~\eqref{the-S}.  In particular, the vacuum-representation
row is
\begin{multline*}
  \Smine_\Omega=\Smine_{13}=
  \Bigl(\!\fffrac{1}{20 \sqrt{5}} ,
  -\fffrac{1}{20 \sqrt{5}} ,
  \fffrac{5-\sqrt{5}+4 \sqrt{10
      (5+\sqrt{5})}}{200} ,
  \fffrac{5-\sqrt{5}-\sqrt{10
      (5+\sqrt{5})}}{200} ,
  \fffrac{5+\sqrt{5}-3 \sqrt{10(5-\sqrt{5})}}{200} ,
  \\
  \fffrac{5+\sqrt{5}+2 \sqrt{10(5-\sqrt{5})}}{200} ,
  \fffrac{-5-\sqrt{5}-2 \sqrt{10(5-\sqrt{5})}}{200} ,
  \fffrac{-5-\sqrt{5}+3 \sqrt{10(5-\sqrt{5})}}{200} ,
  \fffrac{\sqrt{5}-5+\sqrt{10(5+\sqrt{5})}}{200} ,
  \\
  \fffrac{\sqrt{5}-5-4 \sqrt{10 (5+\sqrt{5})}}{200} ,
  \fffrac{1}{10 \sqrt{5}} ,
  -\fffrac{1}{10 \sqrt{5}} ,
  \fffrac{5-\sqrt{5}+4 \sqrt{10(5+\sqrt{5})}}{100} ,
  \fffrac{5-\sqrt{5}-\sqrt{10(5+\sqrt{5})}}{100} ,
  \\
  \fffrac{\sqrt{5}-5-4 \sqrt{10(5+\sqrt{5})}}{100} ,
  \fffrac{\sqrt{5}-5+\sqrt{10(5+\sqrt{5})}}{100} ,
  \fffrac{5+\sqrt{5}-3 \sqrt{10(5-\sqrt{5})}}{100} ,
  \\
  \fffrac{5+\sqrt{5}+2 \sqrt{10(5-\sqrt{5})}}{100} ,
  \fffrac{-5-\sqrt{5}+3 \sqrt{10(5-\sqrt{5})}}{100} ,
  \fffrac{-5-\sqrt{5}-2 \sqrt{10(5-\sqrt{5})}}{100}\Bigr).
\end{multline*}
The matrix $\Kmine$ in~\eqref{The-K} then consists of the blocks
\begin{footnotesize}%
  \begin{multline*}
    \mbox{\normalsize$\Kmine$}{}=
    \mathrm{diag}\Biggl(\!
    20 \sqrt{5},
    -20 \sqrt{5},
    \begin{bmatrix}
      2 \sqrt{2(5-\sqrt{5})} & 1 \\[2pt]
      -2 \sqrt{2(5-\sqrt{5})} &
      \frac{5 \sqrt{5} - 2 + 5 \sqrt{5-2 \sqrt{5}}}{2} 
    \end{bmatrix},
    \\[2pt]
    \begin{bmatrix}
      -2 \sqrt{2
        (5+\sqrt{5})} & 1 \\[2pt]
      2 \sqrt{2
        (5+\sqrt{5})} &
      \frac{109+20 \sqrt{5}-5
        \sqrt{365+158 \sqrt{5}}}{41} 
    \end{bmatrix},
    \begin{bmatrix}
      -2
      \sqrt{2
        (5+\sqrt{5})} & 1 \\[2pt]
      2 \sqrt{2
        (5+\sqrt{5})} &
      \frac{142+15 \sqrt{5}+5
        \sqrt{485+202 \sqrt{5}}}{158}
    \end{bmatrix},
    \\[2pt]
    \begin{bmatrix}
      2 
      \sqrt{2(5 - \sqrt{5})} & 1 \\[2pt]
      -2 \sqrt{2(5 - \sqrt{5})} &
      \frac{379-40 \sqrt{5}-5
        \sqrt{6245-2558
          \sqrt{5}}}{1121}
    \end{bmatrix},
    \begin{bmatrix}
      5 \sqrt{5} & 1 \\[2pt]
      -5 \sqrt{5} & 1
    \end{bmatrix},\qquad\qquad\qquad\qquad
    \\[2pt]
    \begin{bmatrix}
      \sqrt{\frac{5-\sqrt{5}}{2}
      } & 1 &
      1 & \sqrt{\frac{5+\sqrt{5}}{10}
      }
      \\[2pt]          
      -\sqrt{\frac{5-\sqrt{5}}{2}
      } & -1 &
      \frac{5\sqrt{5} - 2 + 5 \sqrt{5-2
          \sqrt{5}}}{2}  & \frac{25
        (\sqrt{5}-1)+\sqrt{
          5450+290 \sqrt{5}}}{20} \\[2pt]
      -\sqrt{\frac{5-\sqrt{5}}{2}
      } & 1 &
      -1 & \sqrt{\frac{5+\sqrt{5}}{10}
      } \\[2pt]
      \sqrt{\frac{5-\sqrt{5}}{2}
      } & -1 &
      \frac{2-5
        \sqrt{5}-5 \sqrt{5-2
          \sqrt{5}}}{2}  & \frac{25
        (\sqrt{5}-1)+\sqrt{
          5450+290 \sqrt{5}}}{20}    
    \end{bmatrix},\qquad\qquad
    \\[2pt]
    \qquad\begin{bmatrix}
      -\sqrt{\frac{5+\sqrt{5}}{2}
      } & 1 &
      1 & -\sqrt{\frac{5-\sqrt{5}}{10}
      } \\[2pt]
      \sqrt{\frac{5+\sqrt{5}}{2}
      } & -1 &
      \frac{109+20 \sqrt{5}-5
        \sqrt{365+158 \sqrt{5}}}{41}&
      \frac{425+125
        \sqrt{5}-\sqrt{476050+79190
          \sqrt{5}}}{410} \\[2pt]
      \sqrt{\frac{5+\sqrt{5}}{2}
      } & 1 &
      -1 & -\sqrt{\frac{5-\sqrt{5}}{10}
      } \\[2pt]          
      -\sqrt{\frac{5+\sqrt{5}}{2}
      } & -1 &
      \frac{-109-20
        \sqrt{5}+5 \sqrt{365+158
          \sqrt{5}}}{41}  &
      \frac{425+125
        \sqrt{5}-\sqrt{476050+79190
          \sqrt{5}}}{410} 
    \end{bmatrix}
    \Biggr).
  \end{multline*}%
\end{footnotesize}%
This gives rise to the fusion-algebra eigenmatrix
$\Pmine=\Smine\Kmine$, shown (at about the limit of reasonable
typesetting capabilities) in Fig.~1.\afterpage{%
  \thispagestyle{empty}
  \rotatebox[origin=lB]{90}{\parbox{.88\textheight}{%
      \mbox{}\kern-150pt {\tiny$\displaystyle
        \setcounter{MaxMatrixCols}{20}
        \renewcommand{\arraycolsep}{-2pt}
    \begin{pmatrix}
      10  & \kern6pt {-10}\kern3pt & 0 & \frac{5+\sqrt{5 (5-2 \sqrt{5})}}{2} & 0 &
      \frac{\sqrt{5 (5-2 \sqrt{5})}}{1+2 \sqrt{5-2 \sqrt{5}}} & 0 &
      \frac{\sqrt{5 (5-2 \sqrt{5})}}{-1+3 \sqrt{5-2 \sqrt{5}}} & 0 &
      \frac{300 \sqrt{5}-40-\sqrt{5 (6245-2558 \sqrt{5})}}{1121} &
      \kern3pt 0\kern3pt
      & \kern3pt\frac{2}{\sqrt{5}}\kern3pt & 0 & 0 & 0 &
      \kern-12pt 5\!-\!\sqrt{5}\!+\!\sqrt{10
        (5\!+\!\sqrt{5})}\;\; & 0 & 0 & 0 &
      \kern-25pt\frac{50+34 \sqrt{5}-4
        \sqrt{1850+110 \sqrt{5}}}{41}
      \\[3pt]
      10 & \kern6pt{-10}\kern3pt & 0 & \frac{5+\sqrt{5 (5-2 \sqrt{5})}}{2} & 0 &
      \frac{\sqrt{5 (5-2 \sqrt{5})}}{1+2 \sqrt{5-2 \sqrt{5}}} & 0 &
      \frac{\sqrt{5 (5-2 \sqrt{5})}}{-1+3 \sqrt{5-2 \sqrt{5}}} & 0 &
      \frac{-40+300 \sqrt{5}-\sqrt{5 (6245-2558 \sqrt{5})}}{1121} & 0
      & -\frac{2}{\sqrt{5}} & 0 & 0 & 0 &
      \kern-16pt \sqrt{5}\!-\!5\!-\!\sqrt{10
        (5\!+\!\sqrt{5})} & 0 & 0 & 0 &
      \kern-24pt\frac{2(-25-17 \sqrt{5}+2
      \sqrt{1850+110 \sqrt{5}})}{41}
    \\[3pt]
      2 & \kern3pt{-2}\kern3pt & 2 & 0 & -2 & 0 & 2 & 0 & -2 & 0 & 0 & \frac{2}{5
        \sqrt{5}} & 0 & \sqrt{2\!+\!\frac{2}{\sqrt{5}}} & 0 & 0 & 0 &
      \sqrt{2\!-\!\frac{2}{\sqrt{5}}} &
      0 & \kern-10pt 0
      \\[3pt]
      8 & \kern3pt{-8}\kern3pt & -2 & \frac{5 \sqrt{5}-5+\sqrt{250-110 \sqrt{5}}}{8} &
      2 & \frac{-170-50 \sqrt{5}+\sqrt{18850+8390 \sqrt{5}}}{164} & -2
      & \frac{-315-75 \sqrt{5}-\sqrt{24650+10910
          \sqrt{5}}}{632}\!\!
      & 2 &
      \!\!\frac{1540-340 \sqrt{5}-\sqrt{315250-139190 \sqrt{5}}}{4484}\; & 0
      & \frac{8}{5 \sqrt{5}} & 0 & \;\;-\sqrt{2\!+\!\frac{2}{\sqrt{5}}} & 0 &
      \kern-16pt \frac{3 \sqrt{5}-5+\sqrt{50-10 \sqrt{5}}}{2}
      & 0 &
      \;\;-\sqrt{2\!-\!\frac{2}{\sqrt{5}}} & 0 &
      \kern-25pt\frac{-55-21 \sqrt{5}+2 \sqrt{3050+1090
          \sqrt{5}}}{41}
      \\[3pt]
      4 & \kern3pt{-4}\kern3pt & \sqrt{5}\!-\!1 & \;\frac{-5-\sqrt{5}-\sqrt{10-2 \sqrt{5}}}{8}\; & \!\!\!1\!+\!\sqrt{5}\!\!\!\! & \frac{10-26
        \sqrt{5}+\sqrt{610+218 \sqrt{5}}}{164} &
      \!\!\!-1\!-\!\sqrt{5}\!\!\! &
      \frac{45-57 \sqrt{5}-\sqrt{890+242 \sqrt{5}}}{632} & 1\!-\!\sqrt{5}
      & \;\frac{\sqrt{11890-2858 \sqrt{5}}-4 (65+73 \sqrt{5})}{4484} & 0
      & -\frac{4}{5 \sqrt{5}}\;\; & 0 & \;\;-\sqrt{2\!-\!\frac{2}{\sqrt{5}}} & 0 &
      \kern-16pt 1\!+\!\sqrt{5\!+\!2 \sqrt{5}}
      & 0 & \sqrt{2\!+\!\frac{2}{\sqrt{5}}} & 0 &
      \kern-25pt\frac{2(2+3 \sqrt{5}-2 \sqrt{125-38
        \sqrt{5}})}{41}
    \\[3pt]
      6 & \kern3pt{-6}\kern5pt & 1\!-\!\sqrt{5} & \frac{-\sqrt{5}-\sqrt{5-2 \sqrt{5}}}{2} &
      \!\!\!-1\!-\!\sqrt{5}\!\!\!\!
      & \frac{30+4 \sqrt{5}-\sqrt{365+158 \sqrt{5}}}{41} &
      \!\!\!1\!+\!\sqrt{5}\!\!\!
      & \frac{60+3 \sqrt{5}+\sqrt{485+202 \sqrt{5}}}{158} &
      \sqrt{5}\!-\!1 & \frac{-300+8 \sqrt{5}+\sqrt{6245-2558
          \sqrt{5}}}{1121} & 0 & -\frac{6}{5 \sqrt{5}}\;\; & 0 &
      \sqrt{2\!-\!\frac{2}{\sqrt{5}}} & 0 &
      \kern-16pt \sqrt{5}\!-\!1\!+\!\sqrt{2 (5\!+\!\sqrt{5})}
      & 0 & \;\;-\sqrt{2\!+\!\frac{2}{\sqrt{5}}} & 0 &
      \kern-25pt-\frac{2(17+5 \sqrt{5}-2 \sqrt{370+22
        \sqrt{5}})}{41}
    \\[3pt]
      6 & \kern3pt{-6}\kern5pt & 1\!-\!\sqrt{5} & \frac{-\sqrt{5}-\sqrt{5-2 \sqrt{5}}}{2} &
      \!\!\!-1\!-\!\sqrt{5}\!\!\!\!
      & \frac{30+4 \sqrt{5}-\sqrt{365+158 \sqrt{5}}}{41} &
      \!\!\!1\!+\!\sqrt{5}\!\!\!
      & \frac{60+3 \sqrt{5}+\sqrt{485+202 \sqrt{5}}}{158} &
      \sqrt{5}\!-\!1 & \frac{-300+8 \sqrt{5}+\sqrt{6245-2558
          \sqrt{5}}}{1121} & 0 & \frac{6}{5 \sqrt{5}} & 0 &
      \;\;-\sqrt{2\!-\!\frac{2}{\sqrt{5}}} & 0 &
      \kern-16pt 1\!-\!\sqrt{5}\!-\!\sqrt{2 (5\!+\!\sqrt{5})}
      & 0 & \sqrt{2\!+\!\frac{2}{\sqrt{5}}} & 0 &
      \kern-25pt\frac{2(17+5 \sqrt{5}-2 \sqrt{370+22
        \sqrt{5}})}{41}
    \\[3pt]
      4 & \kern3pt{-4}\kern3pt & \sqrt{5}\!-\!1 & \;\frac{-5-\sqrt{5}-\sqrt{10-2 \sqrt{5}}}{8}\; &
      \!\!\!1\!+\!\sqrt{5}\!\!\!\!
      & \frac{10-26
        \sqrt{5}+\sqrt{610+218 \sqrt{5}}}{164} &
      \!\!\!-1\!-\!\sqrt{5}\!\!\! &
      \frac{45-57 \sqrt{5}-\sqrt{890+242 \sqrt{5}}}{632} & 1\!-\!\sqrt{5}
      & \;\frac{\sqrt{11890-2858 \sqrt{5}}-4 (65+73 \sqrt{5})}{4484} & 0
      & \frac{4}{5 \sqrt{5}} & 0 & \sqrt{2\!-\!\frac{2}{\sqrt{5}}} & 0 &
      \kern-16pt -1\!-\!\sqrt{5\!+\!2 \sqrt{5}}
      & 0 & \;\;-\sqrt{2\!+\!\frac{2}{\sqrt{5}}} & 0 &
      \kern-25pt-\frac{2(2+3 \sqrt{5}-2 \sqrt{125-38
        \sqrt{5}})}{41}
    \\[3pt]
      8 & \kern3pt{-8}\kern3pt & -2 & \frac{5 \sqrt{5}-5+\sqrt{250-110 \sqrt{5}}}{8} &
      2 & \frac{-170-50 \sqrt{5}+\sqrt{18850+8390 \sqrt{5}}}{164} & -2
      & \frac{-315-75 \sqrt{5}-\sqrt{24650+10910
          \sqrt{5}}}{632}\!\!
      & 2 &
      \!\!\frac{1540-340 \sqrt{5}-\sqrt{315250-139190 \sqrt{5}}}{4484}\; & 0
      & \ -\frac{8}{5 \sqrt{5}}\; \ & 0 & \sqrt{2\!+\!\frac{2}{\sqrt{5}}} & 0 &
      \kern-16pt \frac{5-3 \sqrt{5}-\sqrt{50-10 \sqrt{5}}}{2}
      & 0 &
      \sqrt{2\!-\!\frac{2}{\sqrt{5}}} & 0 &
      \kern-25pt\frac{55+21 \sqrt{5}-2 \sqrt{3050+1090
          \sqrt{5}}}{41}
      \\[3pt]
      2 & \kern3pt{-2}\kern3pt & 2 & 0 & -2 & 0 & 2 & 0 & -2 & 0 & 0 &
      -\frac{2}{5 \sqrt{5}}\;
      & 0 & \;\;-\sqrt{2\!+\!\frac{2}{\sqrt{5}}} & 0 & 0 & 0 &
      \;\;-\sqrt{2\!-\!\frac{2}{\sqrt{5}}} &
      0 & \kern-10pt 0
      \\[3pt]
      5 & \kern3pt 5 & 0 & \frac{5+\sqrt{5 (5-2 \sqrt{5})}}{4} & 0 &
      -\frac{\sqrt{5 (5-2 \sqrt{5})}}{2+4 \sqrt{5-2 \sqrt{5}}} & 0 &
      \frac{\sqrt{5 (5-2 \sqrt{5})}}{-2+6 \sqrt{5-2 \sqrt{5}}} & 0 &
      \frac{40-300 \sqrt{5}+\sqrt{5 (6245-2558 \sqrt{5})}}{2242} & 5 &
      0 & 0 & 0 &
      \kern-10pt 5\!+\!\sqrt{5 (5\!-\!2 \sqrt{5})}\kern-10pt
      & 0 & 0 & 0 &
      -\frac{2 \sqrt{5 (5-2 \sqrt{5})}}{1+2 \sqrt{5-2 \sqrt{5}}}
      & \kern-10pt 0
      \\[3pt]
      5 & \kern3pt 5 & 0 & \frac{5+\sqrt{5 (5-2 \sqrt{5})}}{4} & 0 &
      -\frac{\sqrt{5 (5-2 \sqrt{5})}}{2+4 \sqrt{5-2 \sqrt{5}}} & 0 &
      \frac{\sqrt{5 (5-2 \sqrt{5})}}{-2+6 \sqrt{5-2 \sqrt{5}}} & 0 &
      \frac{40-300 \sqrt{5}+\sqrt{5 (6245-2558 \sqrt{5})}}{2242} & -5
      & 0 & 0 & 0 &
      \kern-10pt-5\!-\!\sqrt{5 (5\!-\!2 \sqrt{5})}\kern-10pt & 0 & 0 & 0 & \frac{2
        \sqrt{5 (5-2 \sqrt{5})}}{1+2
        \sqrt{5-2 \sqrt{5}}} & \kern-10pt0
      \\[3pt]
      1 & \kern3pt 1 & 1 & 0 & 1 & 0 & 1 & 0 & 1 & 0 & 1 & 0 & 1 & 0 & 0 & 0
      & 1 & 0 & 0 & \kern-10pt 0
      \\[3pt]
      4 & \kern3pt 4 & -1 & \frac{5 \sqrt{5}-5+\sqrt{250-110 \sqrt{5}}}{16}
      & -1 & \frac{\sqrt{\frac{5}{2} (5-\sqrt{5})}}{4+8 \sqrt{5-2
          \sqrt{5}}} & -1 &
      \frac{\sqrt{\frac{5}{2}
          (5-\sqrt{5})}}{4-12 \sqrt{5-2 \sqrt{5}}} & -1 &
      \frac{340 \sqrt{5}\!-\!1540\!+\!\sqrt{315250-139190 \sqrt{5}}}{8968} &
      \ 4 \ & 0 & -1 & 0 &
      \kern-10pt\frac{5 \sqrt{5}-5+\sqrt{250-110 \sqrt{5}}}{4}\kern-10pt
      & 0 & -1 & 0 &
      \kern-13pt\frac{170+50 \sqrt{5}-\sqrt{18850+8390 \sqrt{5}}}{82}\kern-13pt
      & \kern-10pt 0
      \\[3pt]
      1 & \kern3pt 1 & 1 & 0 & 1 & 0 & 1 & 0 & 1 & 0 & -1 & 0 & -1 & 0 & 0 &
      0 & -1 & 0 & 0 & \kern-10pt0
      \\[3pt]
      4 & \kern3pt 4 & -1 & \frac{5 \sqrt{5}-5+\sqrt{250-110 \sqrt{5}}}{16} &
      -1 & \frac{\sqrt{\frac{5}{2} (5-\sqrt{5})}}{4+8 \sqrt{5-2
          \sqrt{5}}} & -1 & \frac{\sqrt{\frac{5}{2}
          (5-\sqrt{5})}}{4-12 \sqrt{5-2 \sqrt{5}}} & -1 &
      \frac{340 \sqrt{5}\!-\!1540+\sqrt{315250-139190 \sqrt{5}}}{8968} &
      \ \;-4 \ & 0 & 1 & 0 &
      \kern-10pt\frac{5-5 \sqrt{5}-\sqrt{250-110 \sqrt{5}}}{4}\kern-10pt
      & 0 & 1 & 0 &
      \kern-10pt\frac{-170-50 \sqrt{5}+\sqrt{18850+8390 \sqrt{5}}}{82}\kern-13pt
      & \kern-10pt 0
      \\[3pt]
      2 & \kern3pt 2 & \kern-5pt\frac{\sqrt{5}-1}{2}\kern-5pt & \frac{-5-\sqrt{5}-\sqrt{10-2
          \sqrt{5}}}{16} & \frac{-1\!-\!\sqrt{5}}{2} & \frac{-10+26
        \sqrt{5}-\sqrt{610+218 \sqrt{5}}}{328} & \kern-5pt\frac{-1-\sqrt{5}}{2}\kern-5pt
      & \frac{45-57 \sqrt{5}-\sqrt{890+242 \sqrt{5}}}{1264} &
      \kern-5pt\frac{\sqrt{5}-1}{2}\kern-5pt & \frac{260+292 \sqrt{5}-\sqrt{11890-2858
          \sqrt{5}}}{8968} & -2 & 0 & \kern-5pt\frac{1-\sqrt{5}}{2}\kern-5pt & 0 &
      \kern-10pt\frac{5+\sqrt{5}+\sqrt{10-2 \sqrt{5}}}{4}\kern-10pt
      & 0 &
      \kern-5pt\frac{1+\sqrt{5}}{2}\kern-5pt & 0 &
      \kern-13pt\frac{10-26 \sqrt{5}+\sqrt{610+218 \sqrt{5}}}{82}\kern-13pt
      & \kern-10pt 0
      \\[3pt]
      3 & \kern3pt 3 & \kern-5pt\frac{1-\sqrt{5}}{2}\kern-5pt & \frac{-\sqrt{5}-\sqrt{5-2
          \sqrt{5}}}{4} & \kern-5pt\frac{1+\sqrt{5}}{2}\kern-5pt & \frac{-30-4
        \sqrt{5}+\sqrt{365+158 \sqrt{5}}}{82} & \kern-5pt\frac{1+\sqrt{5}}{2}\kern-5pt &
      \frac{60+3 \sqrt{5}+\sqrt{485+202 \sqrt{5}}}{316} &
      \kern-5pt\frac{1-\sqrt{5}}{2}\kern-5pt & \frac{300-8 \sqrt{5}-\sqrt{6245-2558
          \sqrt{5}}}{2242} & -3 & 0 & \kern-5pt\frac{\sqrt{5}-1}{2}\kern-5pt & 0 &
      \kern-10pt\sqrt{5}\!+\!\sqrt{5-2 \sqrt{5}}\kern-10pt
      & 0 & \kern-5pt\frac{-1-\sqrt{5}}{2}\kern-5pt & 0 &
      \kern-13pt\frac{2(30+4 \sqrt{5}-\sqrt{365+158 \sqrt{5}})}{41}\kern-13pt
    & \kern-10pt 0
    \\[3pt]
      2 & \kern3pt 2 & \kern-5pt\frac{\sqrt{5}-1}{2}\kern-5pt & \frac{-5-\sqrt{5}-\sqrt{10-2
          \sqrt{5}}}{16} & \kern-5pt\frac{-1-\sqrt{5}}{2}\kern-5pt & \frac{-10+26
        \sqrt{5}-\sqrt{610+218 \sqrt{5}}}{328} & \kern-5pt\frac{-1-\sqrt{5}}{2}\kern-5pt
      & \frac{45-57 \sqrt{5}-\sqrt{890+242 \sqrt{5}}}{1264} &
      \kern-5pt\frac{\sqrt{5}-1}{2}\kern-5pt & \frac{260+292 \sqrt{5}-\sqrt{11890-2858
          \sqrt{5}}}{8968} & 2 & 0 & \kern-5pt\frac{\sqrt{5}-1}{2}\kern-5pt & 0 &
      \kern-10pt\frac{-5-\sqrt{5}-\sqrt{10-2 \sqrt{5}}}{4}
      \kern-10pt& 0 &
      \kern-5pt\frac{-1-\sqrt{5}}{2}\kern-5pt & 0 &
      \kern-13pt\frac{-10+26 \sqrt{5}-\sqrt{610+218 \sqrt{5}}}{82}\kern-13pt
      & \kern-10pt 0
      \\[3pt]
      3 & \kern3pt 3 & \kern-5pt\frac{1-\sqrt{5}}{2}\kern-5pt & \frac{-\sqrt{5}-\sqrt{5-2
          \sqrt{5}}}{4} & \kern-5pt\frac{1+\sqrt{5}}{2}\kern-5pt & \frac{-30-4
        \sqrt{5}+\sqrt{365+158 \sqrt{5}}}{82} & \kern-5pt\frac{1+\sqrt{5}}{2}\kern-5pt &
      \frac{60+3 \sqrt{5}+\sqrt{485+202 \sqrt{5}}}{316} &
      \kern-5pt\frac{1-\sqrt{5}}{2}\kern-5pt & \frac{300-8 \sqrt{5}-\sqrt{6245-2558
          \sqrt{5}}}{2242} & 3 & 0 & \kern-5pt\frac{1-\sqrt{5}}{2}\kern-5pt & 0 &
      \kern-10pt-\sqrt{5}\!-\!\sqrt{5\!-\!2 \sqrt{5}}\kern-10pt
      & 0 &
      \kern-5pt\frac{1+\sqrt{5}}{2}\kern-5pt
      & 0 &
      \kern-13pt-\frac{2(30+4 \sqrt{5}-\sqrt{365+158 \sqrt{5}})}{41}\kern-13pt
      & \kern-10pt 0
    \end{pmatrix}$}\\[20pt]
  \small\textsc{Figure}~1. \ \footnotesize The $20\times20$
  eigenmatrix $\Pmine=\Smine\Kmine$ for $\ppi=(5,2)$.  The
  vacuum-representation row is the $13$th.}}\newpage} The
$(p=5,p'=2)$-case of algebra~\eqref{the-fusion} follows from this
$\Pmine$ in accordance with~\eqref{pi-M}--\eqref{MMM}.

\section{Conclusions}
The procedure proposed here is of course not a replacement for the
``honest'' derivation of fusion (cf.~\cite{[EF]}).  We also reiterate
that the success of this procedure is apparently rooted in the quantum
group structure of the corresponding logarithmic conformal field
models~\cite{[FGST3],[FGST-q]} (and actually amounts to no more than
establishing the coincidence with the quantum group Grothendieck
ring).  For the logarithmic $\ppi$ models, anyway, the existence of a
relation between modular transformations of characters and the fusion
additionally supports the ``quantum-group candidate'' for the fusion
of representations of the extended algebra
in~\cite{[FGST3]}.\footnote{In fact, Kazhdan--Lusztig-dual quantum
  groups ``know'' not only about the numerology and modular group
  transformations of extended-algebra characters in logarithmic
  conformal field models but also about the asymptotic form of the
  characters~\cite{[G]}.}  But the much more complicated
``logarithmic'' modular transformations in~\cite{[S-ABC]} are not
likely to yield a fusion algebra similarly.

\subsubsection*{Acknowledgments}
This paper was supported in part by the RFBR Grant 05-01-00996 and by
the Grant~LSS-4401.2006.2.


\begin{thebibliography}{99}
\bibitem{[Gurarie]} V.~Gurarie, \textit{Logarithmic operators in
    conformal field theory}, Nucl.\ Phys.\ B410 (1993) 535
  [hep-th$/$\linebreak[0]9303160].

\bibitem{[GK-rat]} M.R.~Gaberdiel and H.G.~Kausch, \textit{A rational
    logarithmic conformal field theory}, Phys.\ Lett.\ B386 (1996)
  131--137 [hep-th$/$\linebreak[0]9606050].

\bibitem{[G-alg]} M.R.~Gaberdiel, \textit{An algebraic approach to
    logarithmic conformal field theory}, Int.\ J.\ Mod.\ Phys.\ A18
  (2003) 4593--4638 [hep-th$/$\linebreak[0]0111260].

\bibitem{[V]}E.~Verlinde, \textit{Fusion rules and modular
    transformations in 2D conformal field theory}, Nucl. Phys. B~300
  (1988) 360--376.

\bibitem{[V+]} T.~Tsuchiya, K.~Ueno, and T.~Yamada, \textit{Conformal
    field theory on universal family of stable curves with gauge
    symmetries}, Adv. Studies in Pure Math. 19 (1989)
  459--565;
  \\
  E.~Witten, \textit{Two-dimensional gauge theories
    revisited}, J. Geom. Phys. 9 (1992) 303--368;\\
  A.~Bertram and A. Szenes, \textit{Hilbert polynomials of moduli
    spaces of rank $2$ vector bundles. II.} Topology 32 (1993)
  599--609;\\
  Ch.~Sorger, \textit{La formule de Verlinde}, S\'eminaire Bourbaki
  794 (1994) 87--114;\\
  S.~Kumar, M.S.~Narasimhan, and A.~Ramanathan, \textit{Infinite
    grassmannians and moduli spaces of $G$-bundles}, Math. Ann. 300
  (1994) 41--75;\\
  G.~Faltings, \textit{A proof for the Verlinde formula}, J.\ Alg.\
  Geom. 3 (1994) 347--374;\\
  A.~Beauville and Y.~Laszlo, \textit{Conformal blocks and generalized
    theta functions}, Commun. Math. Phys. 164 (1994) 385--419;\\
  G.~Daskalopoulos and R.~Wentworth, \textit{Factorization of rank two
    theta functions. II. Proof of the Verlinde formula},
  Math. Ann. 304 (1996) 21--51;\\
  C.~Teleman, \textit{Verlinde factorization and Lie algebra
    cohomology}, Invent. Math. 126 (1996) 249--263.

\bibitem{[GK-fusion]} M.R.~Gaberdiel and H.G.~Kausch,
  \textit{Indecomposable fusion products}, Nucl.\ Phys. B477 (1996)
  293--318 [hep-th/9604026].

\bibitem{[FK]} M.~Flohr and H.~Knuth, \textit{On Verlinde-like
    formulas in $c_{p,1}$ logarithmic conformal field theories},
  arXiv:0705.0545.

\bibitem{[GR2]} M.R.~Gaberdiel and I.~Runkel, \textit{From boundary to
    bulk in logarithmic CFT}, arXiv:0707.0388 [hep-th].

\bibitem{[FHST]} J.~Fuchs, S.~Hwang, A.M.~Semikhatov, and
  I.Yu.~Tipunin, \textit{Nonsemisimple fusion algebras and the
    Verlinde formula}, Commun.\ Math.\ Phys.\ 247 (2004) 713--742
  [hep-th0306274].

\bibitem{[FGST]} B.L.~Feigin, A.M.~Gainutdinov, A.M.~Semikhatov, and
  I.Yu.~Tipunin, \textit{Modular group representations and fusion in
    logarithmic conformal field theories and in the quantum group
    center}, Commun.\ Math.\ Phys.\ 265 (2006) 47--93
  [hep-th/0504093].

\bibitem{[FGST2]} B.L.~Feigin, A.M.~Gainutdinov, A.M.~Semikhatov, and
  I.Yu.~Tipunin, \textit{Kazhdan--Lusztig correspondence for the
    representation category of the triplet $W$-algebra in logarithmic
    CFT}, Theor.\ Math.\ Phys.\ 148 (2006) 1210--1235
  [math.QA/0512621].

\bibitem{[FGST3]} B.L.~Feigin, A.M.~Gainutdinov, A.M.~Semikhatov, and
  I.Yu.~Tipunin, \textit{Logarithmic extensions of minimal models:
    characters and modular transformations}, Nucl.\ Phys.\ B 757
  (2006) 303--343 [hep-th/0606196].

\bibitem{[FGST-q]} B.L.~Feigin, A.M.~Gainutdinov, A.M.~Semikhatov, and
  I.Yu.~Tipunin, \textit{Kazhdan--Lusztig-dual quantum group for
    logarithmic extensions of Virasoro minimal models}, J.\ Math.\
  Phys.\ 48 (2007) 032303 [math.QA/0606506].

\bibitem{[S-q]} A.M.~Semikhatov, \textit{Factorizable ribbon quantum
    groups in logarithmic conformal field theories}, arXiv:0705.4267
  [hep-th].

\bibitem{[MT]} G.~Mutafyan and I.Yu.~Tipunin, \textit{Double affine
    Hecke algebra in logarithmic conformal field theory},
  arXiv:0707.1625 [math.QA].

\bibitem{[FHT]} P.~Furlan, L.~Hadjiivanov, and I.~Todorov,
  \textit{Zero modes' fusion ring and braid group representations for
    the extended chiral WZNW model}, arXiv:0710.1063 [hep-th].

\bibitem{[KSSB]}T.H.~Koornwinder, B.J.~Schroers, J.K.~Slingerland, and
  F.A.~Bais, \textit{Fourier transform and the Verlinde formula for
    the quantum double of a finite group}, J.\ Phys. A32 (1999)
  8539--8549 [math.QA/9904029].

\bibitem{[Fuchs]} J.~Fuchs, \textit{On non-semisimple fusion rules and
    tensor categories}, hep-th$/$\linebreak[0]0602051.

\bibitem{[EF]} H.~Eberle and M.~Flohr, \textit{Virasoro
    representations and fusion for general augmented minimal models},
  J.\ Phys. A39 (2006) 15245--15286 [hep-th$/$\linebreak[0]0604097].

\bibitem{[G]}A.~Gainutdinov, private communication.
  
\bibitem{[S-ABC]}A.M.~Semikhatov, \textit{Higher string functions,
    higher-level Appell functions, and the logarithmic
    $\widehat{s\ell}(2)_k/u(1)$ CFT model}, arXiv:0710.2028 [math.QA].

\end{thebibliography}
\end{document}